\documentclass[11pt,twoside]{article}

\usepackage{epsfig,amsfonts,color}
\usepackage{amsmath}
\usepackage{mathtools}

\usepackage{amssymb, palatino, geometry,url}
\usepackage{amsthm}
\usepackage{algorithmic}
\usepackage{booktabs}
\usepackage{siunitx}
\usepackage[noresetcount,lined,boxed]{algorithm2e} 

\usepackage[colorlinks=true,linkcolor=blue,citecolor=blue,urlcolor=blue]{hyperref}
\usepackage{subcaption}
\usepackage{cleveref}
\usepackage{float}
\usepackage{fancyhdr}
\usepackage{url}
\usepackage[outdir=./]{epstopdf}
\pdfoutput=1

\theoremstyle{definition}

\newcommand{\be}{\begin{equation}}
\newcommand{\ee}{\end{equation}}
\newcommand{\bea}{\begin{eqnarray}}
\newcommand{\eea}{\end{eqnarray}}

\newcommand{\bvec}{\left(\begin{array}{c}}
\newcommand{\evec}{\end{array}\right)}
\newcommand{\bsub}{\begin{subequations}}
\newcommand{\esub}{\end{subequations}}

\usepackage[thinc]{esdiff}

\usepackage{setspace}
\begin{document}

\title{Automated Characterization and Monitoring of Material Shape using Riemannian Geometry}

\author{Alexander Smith${}^{\text{a}}$\thanks{Corresponding Author: 421 Washington Ave SE \#250, Minneapolis, MN 55455 (smi02527@umn.edu)}, Steven Schilling${}^{\text{b}}$, Prodromos Daoutidis${}^{\text{a}}$\\
{\small ${}^{\text{a}}$Department of Chemical Engineering and Materials Science}\\
{\small \;University of Minnesota, Minneapolis, MN, USA}\\
{\small ${}^{\text{b}}$Technology and Application Development}\\
{\small \;Covia Corp, Independence, OH, USA}\\}

\date{}
\maketitle

\begin{abstract}
Shape affects the physical and chemical properties of a material. Characterizing the roughness, convexity, and general geometry of a material can yield information on its catalytic efficiency, solubility, porosity, and overall effectiveness in the application of interest. However, material shape can be defined in conflicting ways where certain aspects of a material's geometry are emphasized over others, leading to measures of shape that are not generalizable. We explore the use of a Riemannian geometric framework for shape analysis that is generalizable, computationally scalable, and can be integrated into common data analysis methods. Here, material shapes are abstracted as points on a Riemannian manifold. This information can be used to construct statistical moments (e.g., means, variances) and perform tasks such as dimensionality reduction and statistical process control. We provide an introduction to the mathematics of shape analysis and illustrate its application on a manufactured/mined granular material dataset provided by Covia Corp. \\

 \end{abstract}

\section{Introduction}

The geometric shape of a manufactured material impacts both its chemical and physical properties. This is evident in many areas of manufacturing such as 3-dimensional printing, crystal and co-crystal growth, cell morphology in biomanufacturing, and in the production of granular materials such as sand and ceramic microspheres 
 \cite{hossain2015development, ross2016engineering, alshibli2015quantifying, nuchitprasitchai2017factors}. For example, granular materials like sand and ceramic microspheres can be added into liquids such as paints, coatings, and cements to improve hardness, decrease viscosity, increase insulating capability, and reduce the overall amount of volatile organic compounds (VOC's) \cite{suryavanshi2002development, bederina2005reuse}. However, these improvements are directly dependent on the morphology of the individual grains, with spherical grains providing the largest improvement and irregularly shaped particles resulting in decreased performance due to how the particles coalesce at a larger scale \cite{hafid2016effect, altuhafi2013analysis}. This also holds true for granular materials used in metal casting and hydraulic fracturing where the flow of organic materials through the granular medium needs to be closely controlled to prevent metal casting defects and optimal extraction of natural gas and oil from deep rock formations \cite{zheng2016frac, lim2012effects}. Thus, there is a need for methods to automatically quantify and monitor the morphological characteristics of manufactured materials.
Unfortunately, many of the geometric methods employed in manufacturing are application specific. An example is the use of rigid geometric structures for the analysis of crystal morphology in the production of pharmaceuticals. Crystal structure (i.e., crystal form) of an active pharmaceutical ingredient (API) impacts its density, solubility, reactivity, and stability, among other properties \cite{braga2022relevance}. The structure of crystals allows their shape to be quantified with measures such as aspect ratio, form factor, and roundedness, or complex transforms such as wavelet and Fourier transforms \cite{singh2012image, drolon2000particles, bernard1997classification}. However, these methods are difficult to apply to amorphous structures such as those found in cells used in the production of bio-pharmaceuticals, in crystal polymorphisms, or in the mining and refinement of natural materials such as sand \cite{streefland2013process, mercier2014multivariate}. In these cases, the shapes of the materials are often subjectively compared making it difficult to directly quantify differences in their morphology. In the analysis of natural granular material (e.g., sand), a visual chart known as the Krumbein-Sloss chart is used to quantify shape \cite{krumbein1951stratigraphy}. The Krumbein-Sloss chart is used in accordance with the EN:ISO 13503-2:2006 standard, in which 20 grains of sand are randomly chosen and manually compared to illustrations of sand grains to estimate the roundness and sphericity of the particles \cite{ISO13586}. Such methods 
are time-consuming and subjective to the interpretation of the individual performing the analysis. 

Machine learning (ML) methods, such as convolutional neural networks, have recently been proposed as generalizable measures of morphology for materials \cite{wang2021machine, ge2020deep, chowdhury2016image, kim2020machine, smith2020convolutional}. However,  these methods require large amounts of well-sampled training data. In the context of manufacturing and statistical quality control this means that there must be a large number of defective/outlier samples in the training data \cite{weichert2019review}. These data are sparse by nature and may require significant interventions into operating process systems to obtain a diverse enough sample of the various faults that could occur, resulting in significant process downtime and lost revenue. Furthermore, ML methods are difficult to physically interpret and do not provide a direct avenue for robust statistical analysis of their output.

To address these challenges, we propose the use of a mathematical framework to automatically characterize the morphology of  manufactured materials using Riemannian geometry. This framework, often referred to as \emph{statistical shape analysis}, was first introduced by Kendall and defines shape as the geometric information remaining after  rotation, translation, and scaling have been removed from an object  \cite{kendall1977diffusion, kendall2009shape,bookstein1997landmark}. It is based on the realization that geometric shapes can be represented as points on a Riemannian manifold \cite{dryden2016statistical}. The structure of this Riemannian  manifold can be used to directly quantify differences between shapes based on geodesic distances on the manifold. These geodesic distances can be used to develop statistical measures (e.g., means, variances) of a material's intrinsic morphology which can be used in process monitoring and quality control. Furthermore, the Riemannian manifold structure allows us to project data from the manifold onto a tangent (vector) space which can be directly integrated in data analysis tasks such as dimensionality reduction and classification \cite{smith2022data}. This framework has been successfully applied in many research areas such as evolutionary biology \cite{monteiro2000shape, o2000study}, neuroscience \cite{pennec2009statistical, mardia2013alcohol}, chemistry and biochemistry \cite{sauer2003molecular, dryden2007statistical,czogielbayesian}, cell morphology \cite{miolane2021iclr, myers2022regression, gabbert2021septins}, and design of experiments \cite{castillo2011statistical}, but has found little application in the areas of process monitoring and statistical process control.

In this work, we present the mathematical foundations of shape analysis through a Riemannian geometric framework and illustrate its application on a manufactured/mined granular material dataset provided by Covia Corp. We focus on samples of sand and manufactured ceramic microspheres. We analyze microscope images of these samples and develop an automated method for extracting particle shapes from the images. We leverage the presented Riemannian framework to perform dimensionality reduction to visualize the structure of the data and hypothesis testing to understand the morphological differences in the samples. The automated, computationally efficient nature of this framework, coupled with its statistical power,  suggest a powerful statistical process control technique for the continuous improvement and quality control of processes in which shape is a key factor. All code and data needed to reproduce the results presented here can be found at the following link: \url{https://github.com/AdSmithPhd/Shape_Analysis}.

\section{Geometric Shape}

\begin{figure}
     \centering
     \begin{subfigure}[b]{0.24\textwidth}
         \centering
         \includegraphics[width=\textwidth]{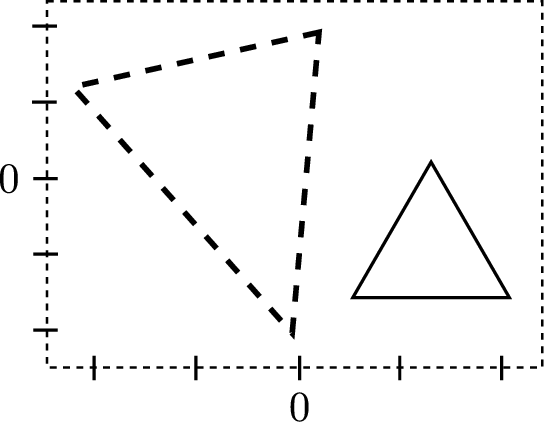}
         \caption{Initialization}
         \label{fig:y equals x}
     \end{subfigure}
     \hfill
     \begin{subfigure}[b]{0.24\textwidth}
         \centering
         \includegraphics[width=\textwidth]{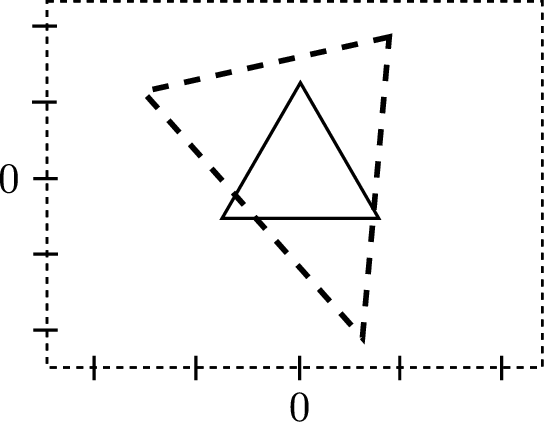}
         \caption{Translation}
         \label{}
     \end{subfigure}
     \hfill
     \begin{subfigure}[b]{0.24\textwidth}
         \centering
         \includegraphics[width=\textwidth]{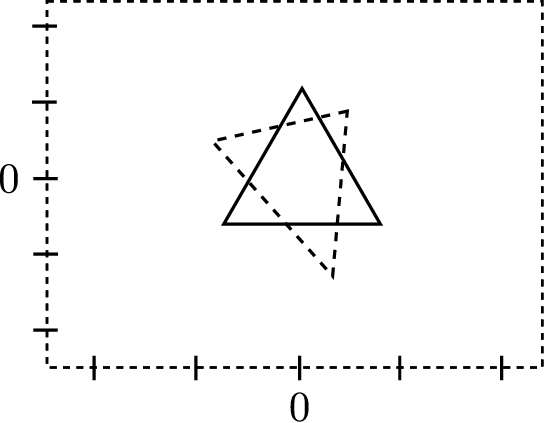}
         \caption{Scaling}
         \label{}
     \end{subfigure}
     \hfill
     \begin{subfigure}[b]{0.24\textwidth}
         \centering
         \includegraphics[width=\textwidth]{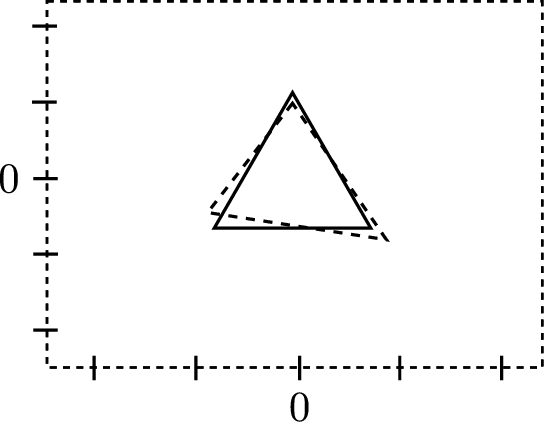}
         \caption{Rotation}
         \label{fig:shap_dif}
     \end{subfigure}
        \caption{Two triangles are translated, scaled, and rotated to emphasize the difference in shape between the two triangles. Once these transformations are completed, the triangles can be analyzed via the Riemannian geometric shape space analysis framework.}
        \label{fig:triangles}
\end{figure}

We first cover the basic mathematics of the analysis of geometric shape. Geometric shape is often defined as the geometric  information that remains when rotation, translation, and scaling are removed from a data object \cite{dryden2016statistical, kendall1977diffusion}. The removal of translation, rotation, and scaling allows us to directly compare the differences in the intrinsic morphology of different objects. For example, Figure \ref{fig:triangles} illustrates an analysis of two triangles in which we translate, scale, and rotate the triangles to best match the other. The remaining geometric differences shown in Figure \ref{fig:shap_dif} represent the difference in \emph{shape} of the triangles.

We will cover the operations of translation, scaling, and rotation individually based on the triangles shown in Figure \ref{fig:triangles} and then extend the methods to general shapes. A key aspect of the analysis of these triangles is that they can be represented by \emph{landmark points}. A landmark point is defined formally as a point of correspondence on each shape that matches between and within populations \cite{dryden2016statistical}. For the triangles in our example the landmark points are given by the location of the corners of each triangle. Thus, each triangle can be represented by a set of $k = 3$ points, ${x_1,x_2,x_3}$, where $x_i \in \mathbb{R}^2$; this is known as the \emph{configuration}. From this configuration we can construct a $3 \times 2$ matrix $\mathbf{X} = [x_1,x_2,x_3]$ that represents the 3 landmark points in 2 dimensions. With this definition we can now understand the mathematics of translation and leverage a centering matrix $\mathbf{C}$ to center our individual triangles around the origin $(0,0)$. The centering matrix is often applied in multivariate statistics and mean centers components of a vector \cite{marden1996analyzing}. It is defined as:

\begin{align}
\mathbf{C} := I_k - \frac{1}{k}J_k
\end{align}
where $I_k$ is the identify matrix of size $k$ and $J_k$ is a $k \times k$ matrix of ones. For example, in our triangular case with three points, $\mathbf{C}_3$ is given as:

\begin{align}
\mathbf{C}_3:=\left[{\begin{array}{rrr}1&0&0\\0&1&0\\0&0&1\end{array}}\right]-{\frac {1}{3}}\left[{\begin{array}{rrr}1&1&1\\1&1&1\\1&1&1\end{array}}\right]=\left[{\begin{array}{rrr}{\frac {2}{3}}&-{\frac {1}{3}}&-{\frac {1}{3}}\\-{\frac {1}{3}}&{\frac {2}{3}}&-{\frac {1}{3}}\\-{\frac {1}{3}}&-{\frac {1}{3}}&{\frac {2}{3}}\end{array}}\right]
\end{align}

We apply $\mathbf{C}$ to the configuration matrices $\mathbf{X}_{C} := \mathbf{C}_3\mathbf{X}$ of both of our triangles in Figure \ref{}. The next step in the shape analysis is to scale our shapes. A direct way to scale a shape is to normalize by the \emph{size} of the shape. Size is a measure that is formally defined as any positive real-valued function of a configuration matrix $f(\mathbf{X}) \rightarrow \mathbb{R}$ such that $f(a\mathbf{X}) = af(\mathbf{X})$ for any $a >0$ and $a \in \mathbb{R}$ \cite{dryden2016statistical}. An often used measure of size is the centroid size, denoted as $S(\mathbf{X}) \rightarrow \mathbb{R}$. The centroid size is defined as \cite{dryden2016statistical}:

\begin{align}
S(\mathbf{X}):= ||\mathbf{C}\mathbf{X}||
\end{align}
where $||\cdot||$ represents the Frobenius (i.e. Euclidean) norm. We can then scale our centered configurations using the centroid size. We will denote these scaled and translated configurations as $\mathbf{Z}$:

\begin{align}
\mathbf{Z} := \frac{\mathbf{X}_C}{||\mathbf{X}_C||}
\end{align}
The scaled and translated versions of our triangles are found in Figure \ref{}. The final operation of rotation is a slightly more difficult operation, and requires an understanding of \emph{pre-shape space}. 

\subsection{Pre-Shape Space and Riemannian Geometry}

\begin{figure}
    \centering
    \includegraphics[width=.4\linewidth]{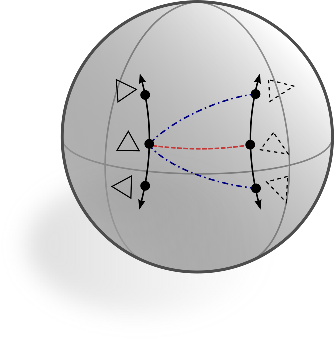}
    \caption{Illustration of the pre-shape Riemannian manifold for triangles. We illustrate geodesics between the solid triangle and three different rotations of the dashed triangle. The dash-dotted, blue geodesics represent geodesics between triangles that have not been optimally rotated. The dashed, red geodesic represents the shortest geodesic between the two optimally aligned triangles. This geodesic quantifies the true difference in shape between the triangles.}
    \label{fig:preshap_man}
\end{figure}

Pre-shape space is an idea first introduced by Kendall and represents the space of all k-point sets (that are not perfectly overlapping) in $\mathbb{R}^m$ that have been properly translated and scaled \cite{kendall1989survey}. This pre-shape space is often denoted as $S_m^k$. For our triangle example in Figure \ref{fig:triangles}, the pre-shape space represents the space of all possible triangular configurations of points and their rotations \cite{dryden2016statistical}. A key notion in the analysis of shapes is that pre-shape space is in fact a hypersphere of unit radius in $(k-1)m-1$ dimensions \cite{kendall1989survey}. Thus, shapes are represented as single points on the surface of this hypersphere. An illustration of the $3$-dimensional pre-shape space sphere for all planar triangles is found in Figure \ref{fig:preshap_man}. 

The pre-shape hypersphere represents a \emph{Riemannian manifold} on which we can measure distance and angles. In general, a manifold $\mathcal{M}$ is a topological space that \emph{locally} resembles a Euclidean space; this means that the neighborhood of a point $p \in \mathcal{U}$ in an $n$-dimensional manifold (with $\mathcal{U} \subseteq \mathcal{M}$) can be mapped to $n$-dimensional Euclidean space through a continuous, bijective function. These neighborhoods and associated mappings are also known as \emph{charts}. For example, the surface of the Earth is a 2D manifold and we can map the curved surface of the Earth to a flat Euclidean plane (i.e., a 2D Euclidean space) using a chart in order to measure properties such as distances or areas. A \emph{Riemannian} manifold represents a differentiable manifold $\mathcal{M}$ on which there exists a defined tangent space $T_pM$ at each point $p \in \mathcal{M}$ along with a defined inner product $g_p: T_pM \times T_pM \rightarrow \mathbb{R}$. This inner product allows us to compute distances and angles between points on the manifold with respect to the geometry of the Riemannian manifold. A rigorous introduction to the mathematics of Riemannian manifolds can be found in our previous work and other resources \cite{smith2022data, pennec2006intrinsic, guigui2023introduction}.

Pre-shape space represents a simple Riemannian manifold, the unit sphere \cite{kendall1989survey, dryden2016statistical}. For example, the distance between two points on a unit sphere is exactly equal to the angle between the points. Thus, for our analysis we can compute the distance between two points on the pre-shape space sphere through an inner product. For two points (e.g., shapes) $\mathbf{Z}_i, \mathbf{Z}_j \in  S_m^k$ we can compute their Frobenius inner product as:

\begin{align}
\langle \mathbf{Z}_i,\mathbf{Z}_j\rangle := \text{tr}(\mathbf{Z}_i^T\mathbf{Z}_j) = \sum_{i=1}^m \lambda_i
\end{align}
where $\lambda_i$ represent the eigenvalues of the matrix $\mathbf{Z}_i^T\mathbf{Z}_j$. We can convert this inner product into an angle, and thus a distance on the pre-shape space. We denote this distance as a \emph{geodesic} distance $\rho(\mathbf{Z}_i,\mathbf{Z}_j)$ defined as:

\begin{align}
\rho(\mathbf{Z}_i,\mathbf{Z}_j) := \text{arccos}(\sum_{i=1}^m \lambda_i)
\end{align}

An illustration of the geodesic distance between two points on the pre-shape space sphere is found in Figure \ref{fig:preshap_man} represented by the red, dashed line and blue, dash-dotted lines. The Riemannian geometry of the pre-shape space provides a concise mathematical framework for the analysis of shape. However, the pre-shape space does not account for rotation of shapes. Thus, our previously defined geodesic distances on the pre-shape space sphere may compute a difference between two shapes that reflects a difference in rotation and shape rather than shape alone. This is illustrated in Figure \ref{fig:preshap_man}, where the blue, dash-dotted lines represent geodesics on the pre-shape space between triangles that are not optimally rotated. The red, dashed line represents the smallest geodesic distance between the two shapes, capturing the true difference in shape of the triangles with optimal rotation. 

Thus, there is a need to optimize the rotation of two shapes such that there is optimal alignment on the pre-shape space and we are able to compute a geodesic between the shapes that reflects changes in shape rather than rotation. We can frame this as an optimization problem and use this framework to identify an optimal \emph{rotation matrix} for one of the shapes. A rotation matrix is formalized as an $m \times m$ matrix denoted as $\Gamma$ where $\Gamma^T\Gamma = \Gamma\Gamma^T = I_m$ and $\text{det}(\Gamma) = +1$.   $I_m$ is the identity matrix of size $m$ and $\text{det}(\cdot)$ represents the matrix determinant \cite{dryden2016statistical}. The set of all $m \times m$ rotation matrices is formally known as the special orthogonal group denoted SO(m). A simple method for computing an optimal rotation (alignment) between two shapes $\mathbf{X}_i, \mathbf{X}_j$ is through the following formulation, which is known as the \emph{partial Procrustes distance} denoted as $d_p(\cdot)$ \cite{rohlf1999shape}:

\begin{align}
d_p(\mathbf{X_i},\mathbf{X_j}) := \inf_{\mathbf{\Gamma} \in SO(m)} \ ||\mathbf{Z}_i - \mathbf{Z}_j\Gamma||
\end{align}
where $||\cdot||$ represents the Frobenius norm and $\mathbf{Z}_i, \mathbf{Z}_j \in  S_m^k$ represent the shapes $\mathbf{X_i},\mathbf{X_j}$ mapped to the pre-shape space. This can be rewritten analytically as:

\begin{align}
d_p(\mathbf{X_1},\mathbf{X_2}) = \sqrt{2}\left(1 - \sum_{i=1}^m \lambda_i \right)^{1/2}
\end{align}
where $\lambda_i$ represent the square roots of the eigenvalues of $\mathbf{Z}_1^T\mathbf{Z}_2\mathbf{Z}_2^T\mathbf{Z}_1$. The optimal rotation is given as $\hat\Gamma$:

\begin{align}
\hat\Gamma := \mathbf{U}\mathbf{V}^T
\end{align} 
where $\mathbf{U}$ and $\mathbf{V}$ are from the singular value decomposition of $\mathbf{Z}_2^T\mathbf{Z}_1$:

\begin{align}
\mathbf{Z}_2^T\mathbf{Z}_1 = \mathbf{V}\mathbf{\Lambda}\mathbf{U}^T
\end{align}

Using this information, we can formulate the geodesic distance between two shapes that have been optimally rotated. We denote this distance as $\hat{\rho}(\mathbf{X}_i,\mathbf{X}_j)$ and define it as:

\begin{align}
\hat{\rho}(\mathbf{X}_i,\mathbf{X}_j) := 2 \ \text{arcsin}(d_p(\mathbf{X_1},\mathbf{X_2})/2) 
\end{align}

The optimal geodesic distance now represents a measure of the true geometric difference between two shapes. Formally, this distance now represents a proper geodesic distance on the \emph{shape space}. The shape space represents the space of all k-point sets (that are not perfectly overlapping) in $\mathbb{R}^m$ that have been properly translated, scaled, \emph{and  rotated} \cite{dryden2016statistical, kendall1989survey}. The shape space contains all information about a given configuration matrix $\mathbf{X}$ that is invariant under translation, rotation and scaling. This is often denoted by the set:

\begin{align}
[\mathbf{X}] := \{\mathbf{Z}\Gamma : \Gamma \in SO(m)\}
\end{align}

This means that a given pre-shape $\mathbf{Z} \in S_m^k$ and all its possible rotations $\{\mathbf{Z}\Gamma : \Gamma \in SO(m)\}$ are mapped to single point in shape space. Thus, we are able to now leverage the proper geodesics in the shape space to compare the geometric shape difference between objects. However, one limiting factor in our analysis is that the computed geodesic distances represent non-linear relationships between different shapes. This makes it difficult to directly apply many data-centric techniques that assume that data lies in a linear (vector) space. One way to address this issue is to leverage the tangent space of the pre-shape space manifold, which is a linear (vector) space, and project our data to this linear space which retains much of the geometric information of the pre-shape manifold. 

\subsection{Tangent Space Analysis}

\begin{figure}[htp!]
     \centering
     \begin{subfigure}[b]{0.45\textwidth}
         \centering
         \includegraphics[width=\textwidth]{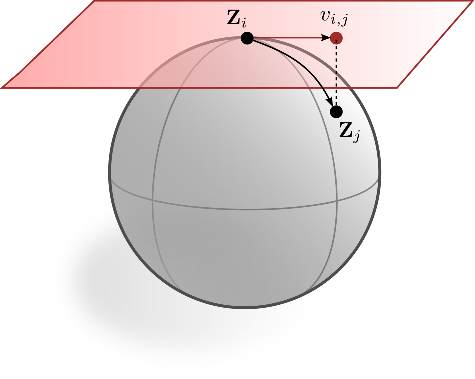}
         \caption{Tangent space at point $\mathbf{Z}_i \in S_m^k$}
         \label{fig:y equals x}
     \end{subfigure}
     \hfill
     \begin{subfigure}[b]{0.35\textwidth}
         \centering
         \includegraphics[width=\textwidth]{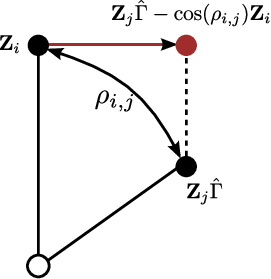}
         \caption{Tangent space projection}
         \label{fig:tan_proj}
     \end{subfigure}
     \caption{(a) Illustration of the tangent space to a point $\mathbf{Z}_i \in S_m^k$ with the tangent vector $v_{i,j}$ representing the projection of the geodesic between $\mathbf{Z}_i, \mathbf{Z}_i \in S_m^k$ to the tangent space. (b) Illustration of how a point $\mathbf{Z}_j \in S_m^k$ can be mapped to the horizontal tangent space centered at $\mathbf{Z}_i \in S_m^k$. To project to the horizontal component of the tangent space, rather than the full tangent space, we must apply optimal rotation matrix $\hat{\Gamma}$ to $\mathbf{Z}_j$ prior to the tangent space projection.}
    
\label{fig:tanspace}
\end{figure}

For a manifold $\mathcal{M}$, such as the pre-shape space, the \emph{tangent space} $T_p M$ at a point $p \in \mathcal{M}$ is formally defined as the space of derivates of all \emph{curves} passing through $p$ \cite{smith2022data}. A curve is defined as a continuous function $\gamma : [a,b] \rightarrow \mathcal{M}$ which maps the interval $[a,b] \in \mathbb{R}$ to the manifold $\mathcal{M}$. Thus, the tangent space $T_p M$ is built from all vectors $v = \dot{\gamma(0)} \in T_pM$ where $\gamma(0) = p \in \mathcal{M}$. The tangent space at a point $p \in \mathcal{M}$ represents a linear vector space that is of the same dimension as the manifold from which it is constructed. A visualization of the tangent space for the triangle pre-shape sphere is found in Figure \ref{fig:tanspace}. 

Another key aspect in the analysis of tangent spaces is the mapping from the tangent space to the manifold, known as the \emph{exponential map} and inversely, the \emph{logarithmic map}. For a tangent space defined at a point $p \in \mathcal{M}$ we can construct a curve $\gamma: [0,1] \rightarrow \mathcal{M}$ such that $\gamma(0) = p$, $\dot\gamma(0) = v \in T_pM$ and $\gamma(1) = q \in \mathcal{M}$. We define the exponential map as the mapping from the vector $v \in T_pM$ to the endpoint of the curve $q \in \mathcal{M}$:

\begin{align}
\text{exp}_p(v) = q
\end{align}
We  define the inverse of the exponential map as the logarithmic map, which maps the point $q \in \mathcal{M}$ to the vector $v \in T_pM$:

\begin{align}
\text{log}_p(q) = v
\end{align}

These operations can be defined for any constructed tangent space on the surface of a Riemannian manifold and are illustrated in Figure \ref{fig:tanspace}. For our pre-shape space sphere we will focus on an analytic formulation for the logarithmic mapping as we are interested in mapping our data from the non-linear surface of the sphere to the linear tangent space. There are many choices for construction of a tangent space for the pre-shape sphere, but we will focus on the \emph{Procrustes tangent space} as it is interpretable and has proven effective in many application areas \cite{dryden2016statistical}. For two shapes on the pre-shape manifold $\mathbf{Z}_i,\mathbf{Z}_j \in S_m^k$ we can construct a tangent space $T_{Z_{i}}M$ at $\mathbf{Z}_i$ and project $\mathbf{Z}_j$ to the Procustes tangent space as:

\begin{align}
\hat{v}_{i,j} := \mathbf{Z}_j\hat{\Gamma} - \text{cos}(\rho_{i,j})\mathbf{Z}_i
\end{align}
where $\rho_{i,j}$ represents the geodesic distance between $\mathbf{Z}_i$ and $\mathbf{Z}_j$. The Procrustes tangent space differs from the full tangent space at the point $\mathbf{Z}_i \in S_m^k$ as it accounts for the rotation of the pre-shape $\mathbf{Z}_j \in S_m^k$ by applying the optimal rotation matrix $\hat\Gamma$ derived from Procrustes analysis. This distinction is often denoted by a decomposition of the tangent space of $S_m^k$ into a \emph{horizontal} component that captures changes in shape and a \emph{vertical} component that accounts for changes in rotation \cite{rohlf1999shape, dryden2016statistical}. Thus, we are focused on the horizontal component of the tangent space as we are interested in differences with respect to shape rather than rotation, which is illustrated in Figure \ref{fig:tan_proj}. Through this projection, we are able to project the non-linear structure of the pre-shape manifold into a linear (vector) space. This linear space can then be used to construct hypothesis tests, execute statistical process control methods, and perform common data centric tasks such as dimensionality reduction and regression/classification. 

\section{Application: Granular Material Morphology}

\begin{figure}
     \centering
     \begin{subfigure}[b]{0.3\textwidth}
         \centering
         \scalebox{1}[-1]{\includegraphics[width=\textwidth]{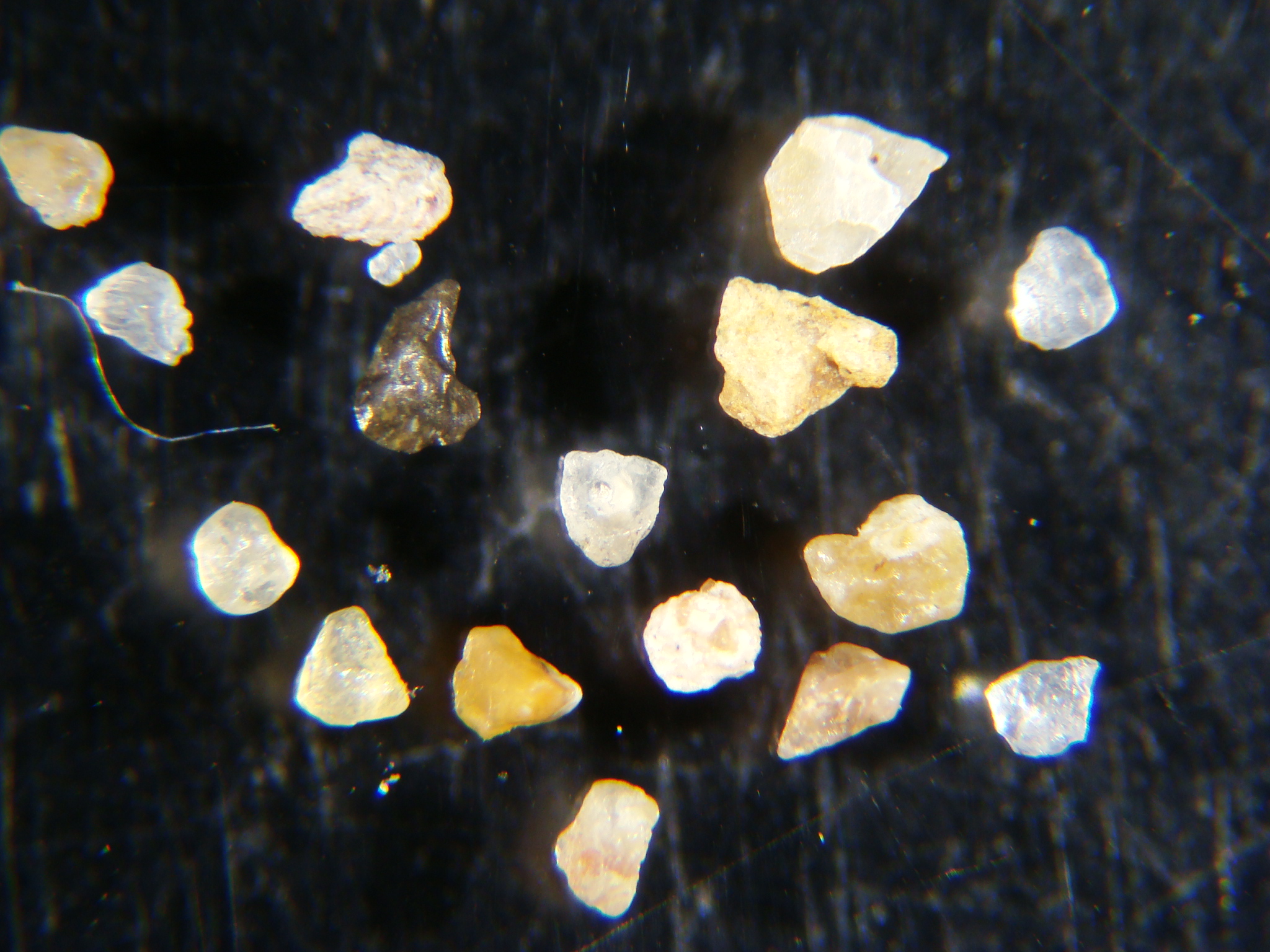}}
         \caption{Angular Sand}
         \label{fig:y equals x}
     \end{subfigure}
     \hfill
     \begin{subfigure}[b]{0.3\textwidth}
         \centering
         \includegraphics[width=\textwidth]{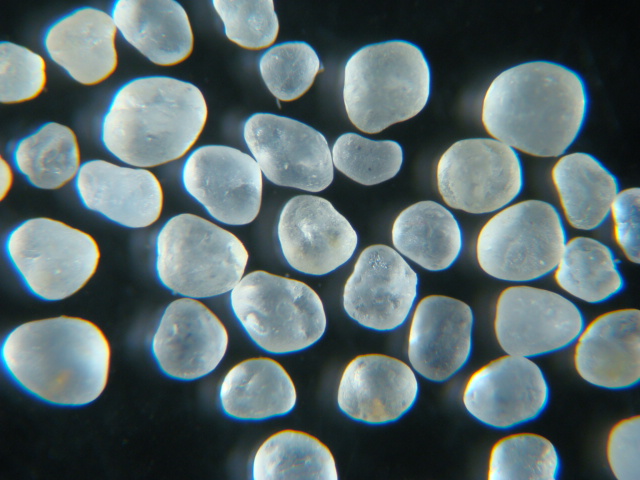}
         \caption{Round Sand}
         \label{}
     \end{subfigure}
     \hfill
     \begin{subfigure}[b]{0.3\textwidth}
         \centering
         \includegraphics[width=\textwidth]{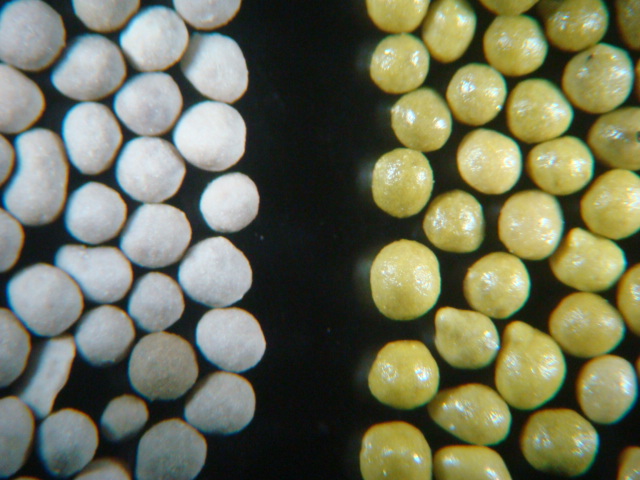}
         \caption{Ceramic Microspheres}
         \label{}
     \end{subfigure}
        \caption{Examples of sands mined at different geographic locations (a-b) and manufactured ceramic microspheres (c).}
        \label{fig:samples}
\end{figure}

We now focus on an application of the geometry of shape space in the analysis of granular materials such as sand and ceramic microspheres. Granular material particle geometry influences the function and bulk physical properties of many manufactured materials. This holds true in many areas of manufacturing such as pharmaceuticals, paints, coatings, metal casting, 3-dimensional printing, and hydraulic fracturing, among others \cite{suryavanshi2002development, hafid2016effect, braga2022relevance, streefland2013process}. A significant challenge in the analysis of particle geometry is the characterization of particles that are amorphous in nature and difficult to quantify using simple geometric descriptors such as aspect ratios or sphericity. This is especially problematic in the manufacture/mining of sand and production of ceramic microspheres where geometry can fluctuate in ways that are not easily modeled. Figure \ref{fig:samples} shows example images from Covia corp of sand mined in different geographic locations that exhibit angular and round geometry and ceramic microsphere samples. 

Figure \ref{fig:samples} illustrates the multitude of ways in which the geometry of particles can fluctuate, demonstrating the difficulty in producing a framework for quantifying the quality and variability of particle morphology. Currently, the main method for characterizing the morphology of particle samples is through manual comparison with the Krumbien-Sloss chart, which quantifies the sphericity of a particle with its roundness (e.g., smoothness) \cite{ISO13586, krumbein1951stratigraphy}. The Krumbien-Sloss chart contains multiple illustrations of particles that are compared against physical particles to obtain an estimate of both roundess and sphericity. Particles that are both highly smooth and round are considered high-quality material as the bulk physical properties of the material will be homogeneous and predictable. Multiple problems exist with the Krumbien-Sloss method. First, the method is manual, making it time consuming and dependent on the individual performing the analysis. Second, there are multiple Krumbien-Sloss charts, making the resulting quantification of the sample dependent on the chart used. We will leverage the tools of shape space analysis to automatically quantify the morphology of the samples shown in Figure \ref{fig:samples} without the use of a Krumbien-Sloss chart. We will also demonstrate how shape space analysis can be used to develop statistical tests for changes in particle morphology and for the implementation of dimensionality reduction techniques such as principal component analysis (PCA).

\subsection{Data Pre-Processing}

\begin{figure}
     \centering
     \begin{subfigure}[b]{0.32\textwidth}
         \centering
         \includegraphics[width=\textwidth]{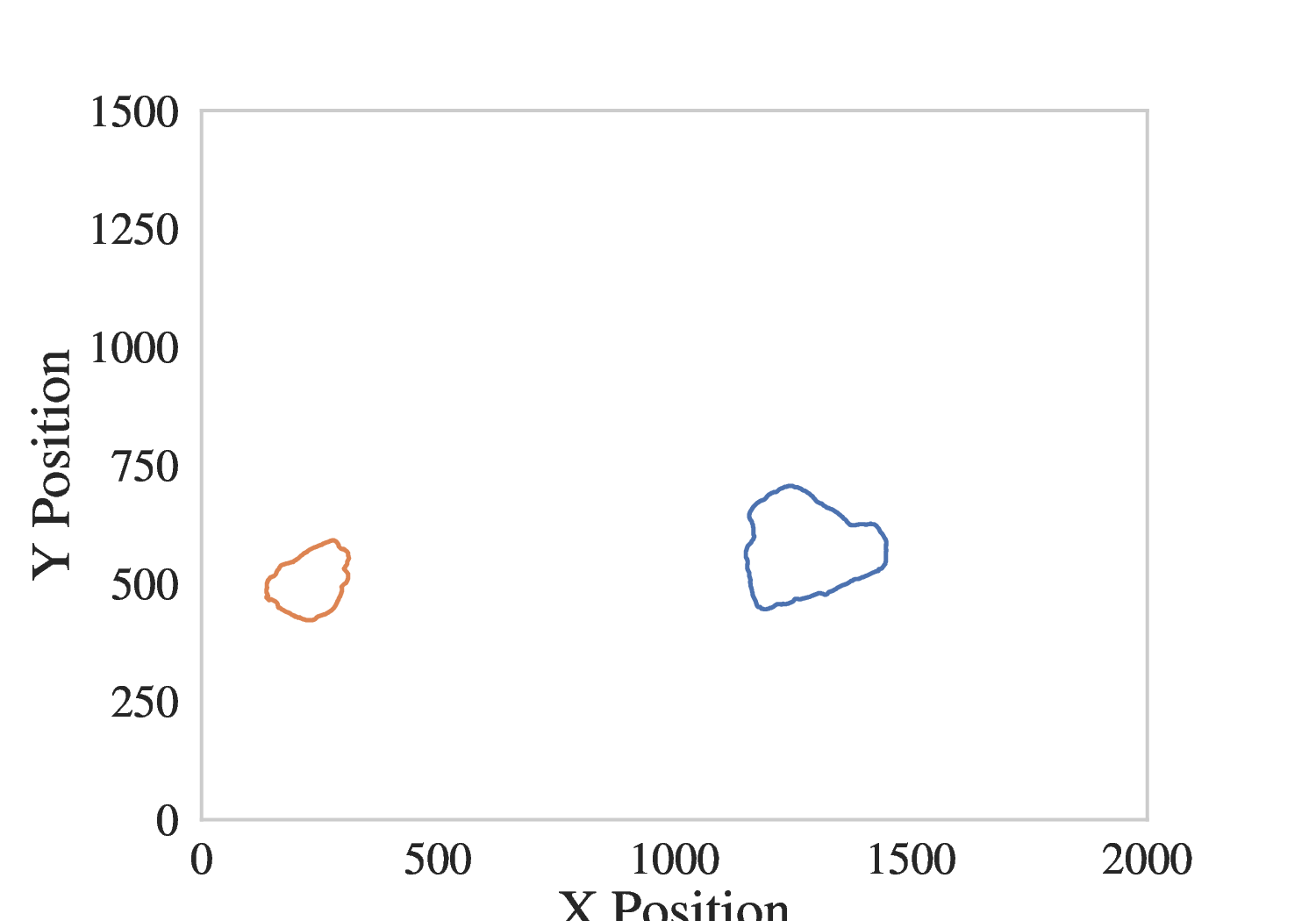}
         \caption{Original contours}
         \label{fig:okay}
     \end{subfigure}
     \hfill
     \begin{subfigure}[b]{0.32\textwidth}
         \centering
         \includegraphics[width=\textwidth]{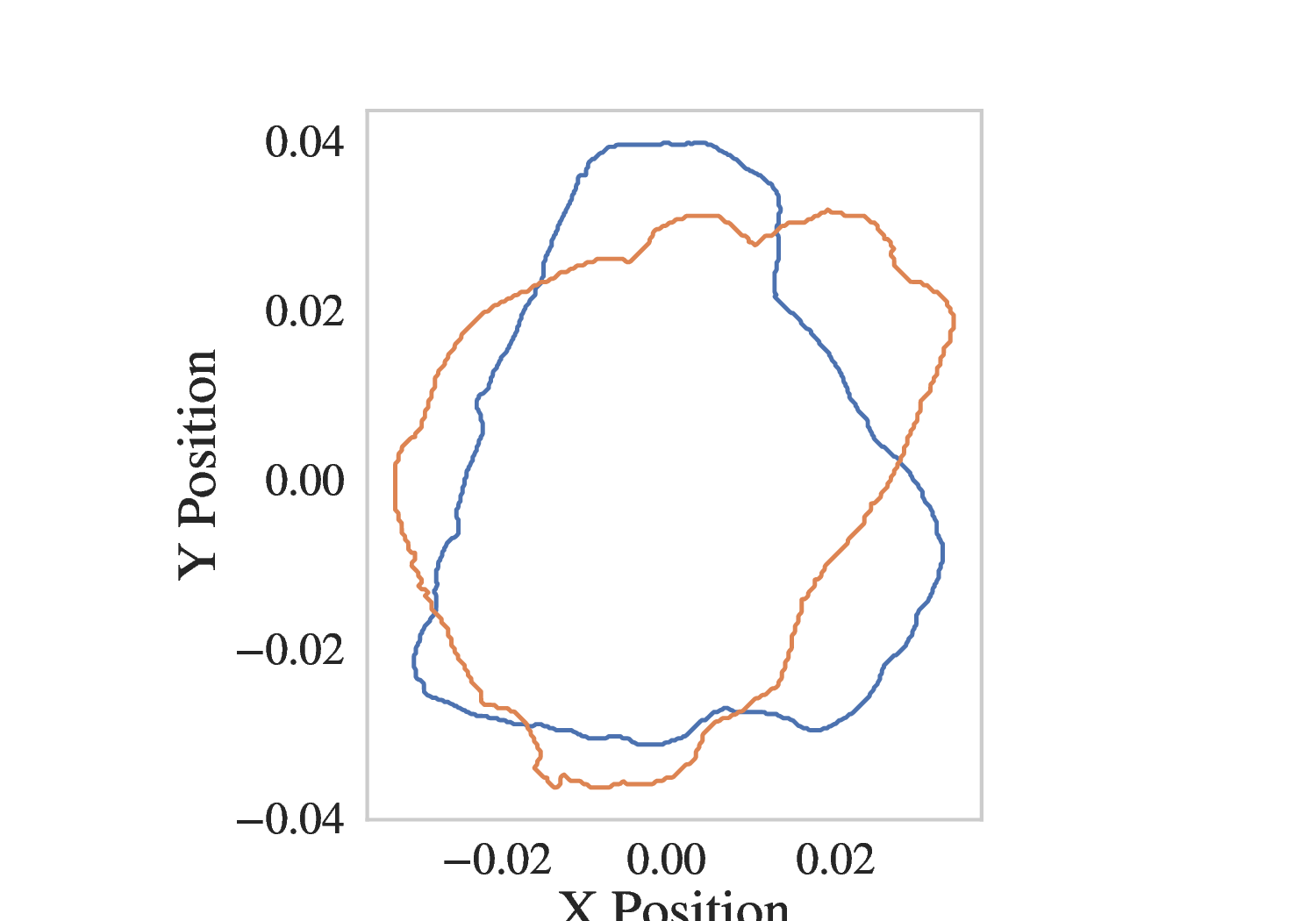}
         \caption{Translated and scaled}
         \label{}
     \end{subfigure}
     \hfill
     \begin{subfigure}[b]{0.32\textwidth}
         \centering
         \includegraphics[width=\textwidth]{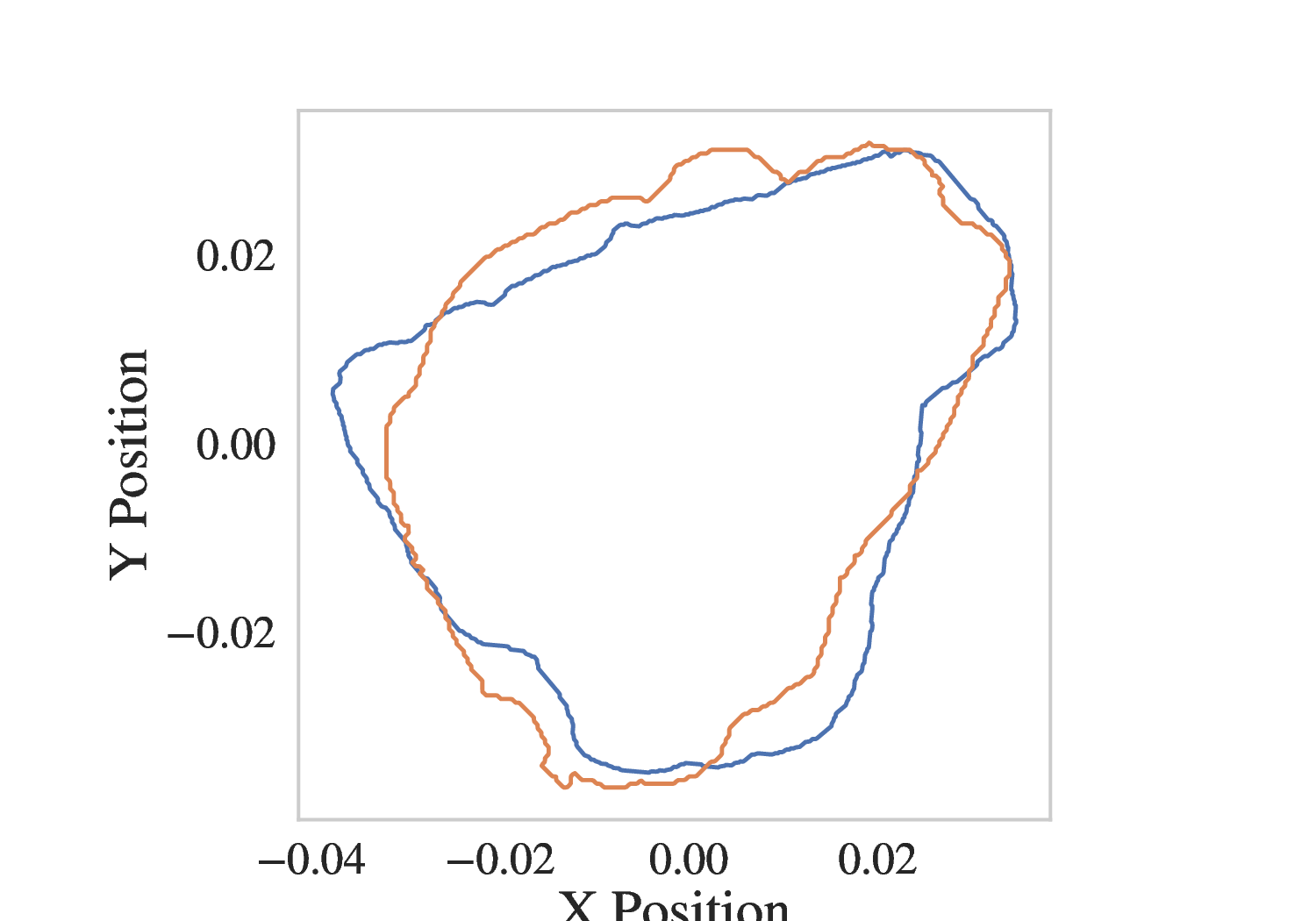}
         \caption{Reparameterized and rotated}
         \label{}
     \end{subfigure}
        \caption{Illustration of the translation, scaling, and optimal rotation of two angular sand outlines.}
        \label{fig:outlines}
\end{figure}

We first take each of our images and extract an outline for each individual particle. This is done through a common method using watershed segmentation in the \texttt{OpenCV} package implemented in \texttt{Python}. We note in our analysis that these images contain particles that are slightly overlapping or touching. This makes it difficult to extract a perfect outline for each individual particle, thus, we focus on particles in which we can extract a full outline. To compare each particle using shape space analysis, we must first interpolate/resample the outline of each particle so that each outline contains the same number of points ensuring they are on the same pre-shape manifold. Two resampled/interpolated outlines from the image in Figure \ref{fig:y equals x} are shown in Figure \ref{fig:okay}. Once the data is pre-processed we can translate, rotate, and scale each particle outline, shown in Figure \ref{fig:outlines}, so that we can measure differences in their shape. 

One difficulty faced in this analysis is that we are dealing with shapes that do not have defined landmark points. Landmark points represent points on a shape that can be consistently identified even when a shape is deformed. For example, the triangle has three obvious landmarks which are the vertices of the three corners of the triangle. Thus, for our data we must consider translation, rotation, scaling, and reparameterization of the shape. This has been deeply explored in the current literature, and one class of methods that are highly effective in the analysis of data with no-defined landmarks are the so-called \emph{elastic metrics} initially proposed in the work of Srivastava and co-workers \cite{srivastava2010shape}. We will not cover this method in detail, but the elastic metric proposed by Srivastava and co-workers represents parameterized curves as points on a spherical pre-shape manifold through the square root velocity function (SRVF). Given a real-valued curve $f(t): [0,1] \rightarrow \mathbb{R}^m$ the SRVF is given as $q: [0,1] \rightarrow \mathbb{R}^m$, which is defined as:

\begin{align}
q(t) := \frac{\dot{f(t)}}{\sqrt{||\dot{f(t)}||}}
\end{align} 
where $||\cdot||$ represents the Euclidean norm and $\dot{f(t)}$ represents the derivative of $f$ with respect to $t$. The SRVF $q(t)$ is invariant to translation, and can be further transformed to be invariant to scaling, rotation, and reparameterization \cite{srivastava2010shape, le2019discrete}. Similar to our pre-shape space for configuration matrices, this representation and its associated pre-shape manifold can be used to construct geodesic distances and projections to a tangent space, allowing us to perform statistical analysis and other data-centric tasks like dimensionality reduction which we demonstrate in the following section. To accomplish this we leverage the powerful \texttt{Geomstats} Python package and its computational implementation of the elastic metric within a discrete framework developed by Brigant \cite{le2019discrete, miolane2020geomstats}.

\subsection{Data Analysis}

\begin{figure}[htp!]
    \centering
    \includegraphics[width=.55\linewidth]{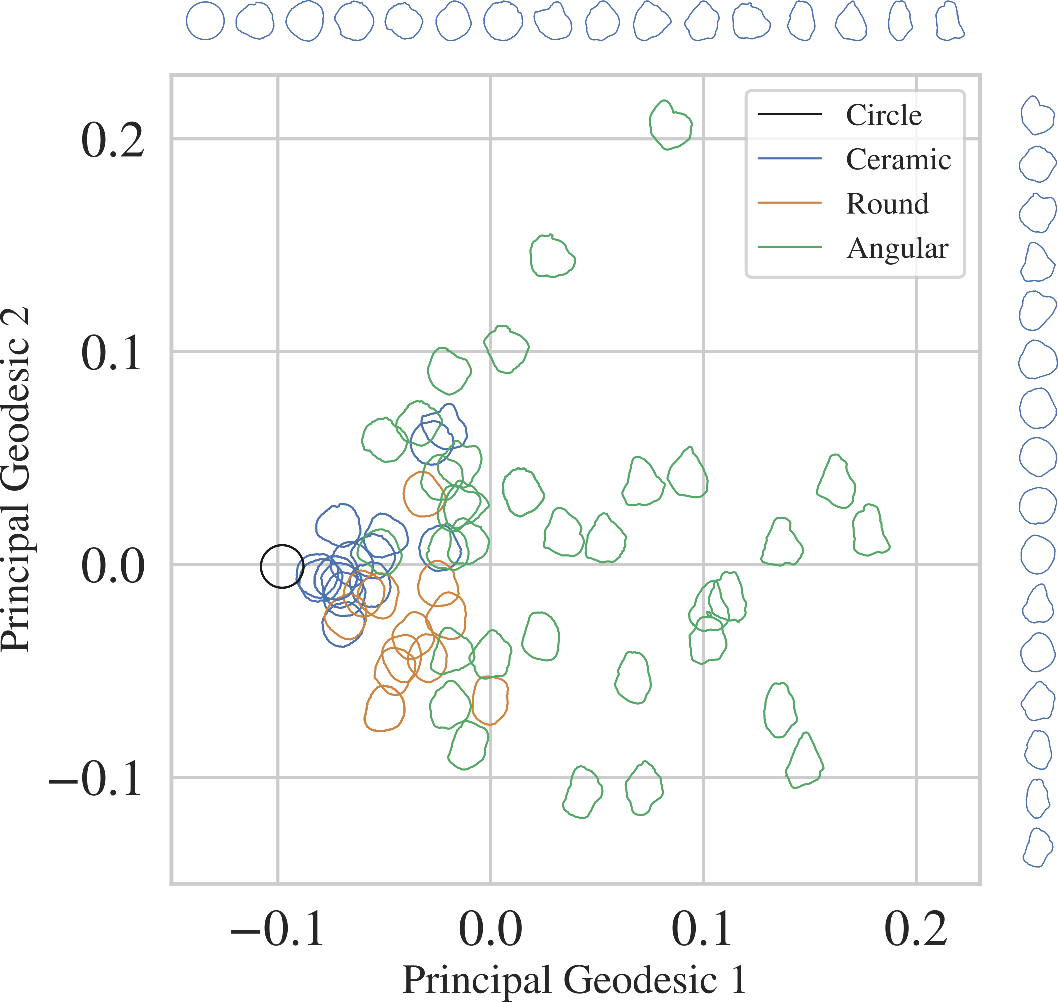}
    \caption{Principal geodesic analysis of the particles outlines from angular sand, round sand, and ceramic microsphere samples. The base point of defined tangent space is the perfect circle. We project to the first two principal geodesics that capture the most variance in the data with respect to the shape space of our data. The projection illustrates the relative geodesic distances to the perfect sphere for the three samples. The round sand and ceramic microspheres have the tightest distribution around the perfect circle. We also illustrate on the edges of the image the changes in morphology along the two principal geodesics.}
    \label{fig:text_pca}
\end{figure}

Once our data has been properly pre-processed, we can  leverage the framework introduced in the previous sections to peform both dimensionality reduction and statistical analysis for our particle shapes. In our analysis, we are interested in comparing our particles to a perfect circle, which represents a particle with perfect sphericity and roundness. The geodesic distance from the sphere provides a direct representation of the quality of the particles taken from the sample. As can be seen in Figure \ref{} we should expect the ceramic microspheres and the round sand samples to be geodesically close to the perfect sphere with minimal variance in the distance and the angular sand sample to be much farther away. First, we visualize the morphological differences in the samples by performing principal component analysis (PCA) on the horizontal tangent space of the pre-shape space manifold centered at the circle. This form of PCA, introduced by Fletcher and co-workers, is often referred to as principal geodesic analysis (PGA) as the method identifies geodesic basis with respect to the manifold (via the tangent space) rather than basis vectors in Euclidean space \cite{fletcher2004principal}. Figure \ref{fig:text_pca} shows the first two principal geodesics of PGA performed on the optimally aligned particle shapes. We can see that the analysis reflects the intuition that the manufactured ceramic microspheres are highly concentrated near the circle, with the round sand being slightly farther from the circle, and the angular sand being of a greater distance with a much larger variation in the distance. 

\begin{figure}[htp!]
    \centering
    \includegraphics[width=.7\linewidth]{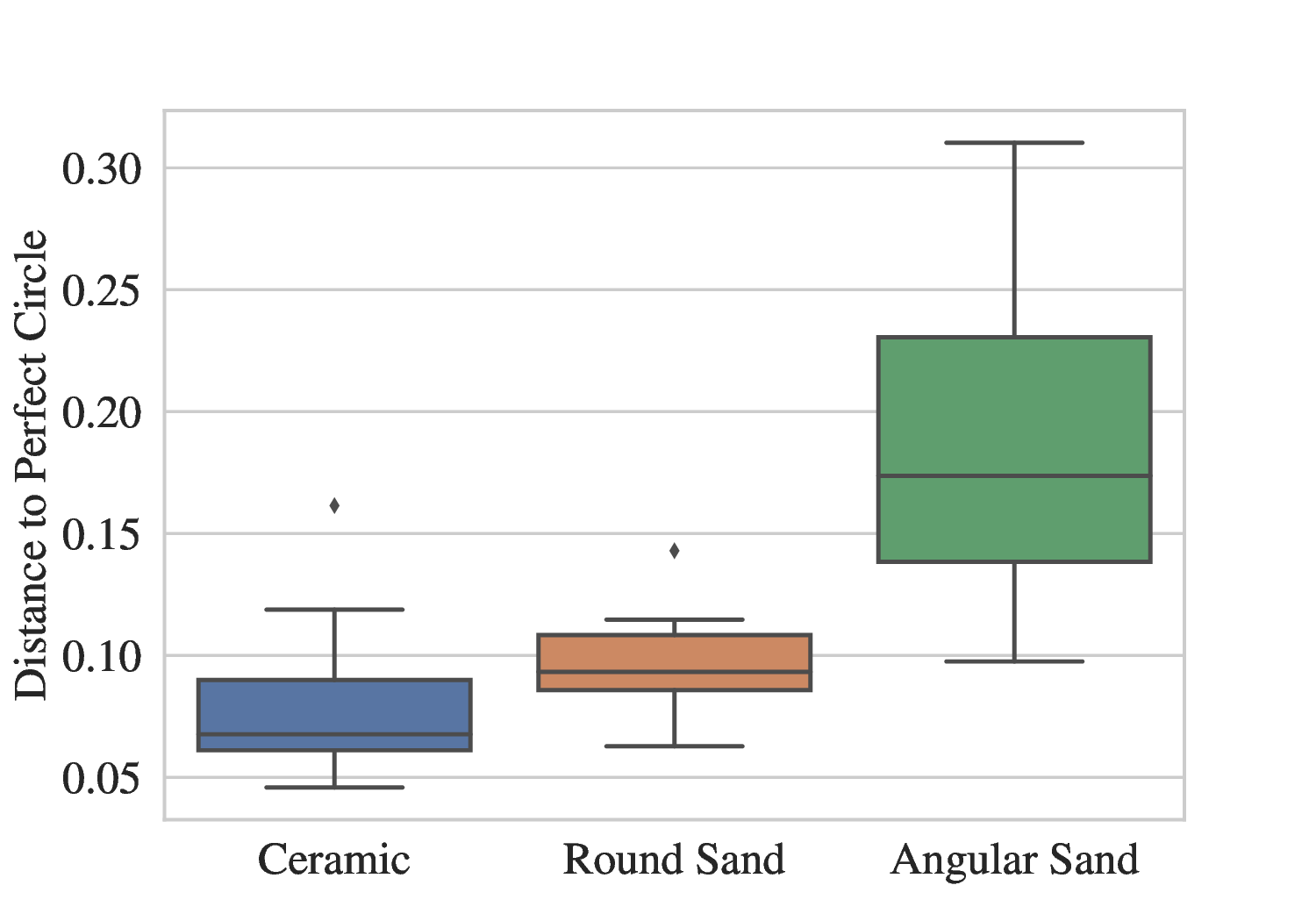}
    \caption{Box plots of the tangent space distance between each particle outline and the perfect circle. We can see that the synthetic ceramic microspheres have the tightest morphological distribution around the sphere, with the round sound being slighty farther away, and the angular sand have the largest mean distance and largest variation in morphology.}
    \label{fig:box}
\end{figure}

We can leverage these results to perform statistical analysis and understand how the distance in the tangent space can be used to distinguish between the samples, and  whether or not there is a statistically significant difference between each of the particle samples. Figure \ref{fig:box} shows box plots of the distances in the tangent space from each particle to the perfect circle for our samples. From the analysis, we see that both the mean and variance of the geodesic distance from the circle for the ceramic microspheres and round sand is much lower than the angular sand. To quantify this difference we can perform a 2-sample t-test between the three different samples. The results of the 2-sample t-test are shown in Table \ref{tbl:1}. We find that there is a statistically significant difference in shape between the ceramic, the round sand and the angular sand samples, with an $\alpha = 0.05$. Figure \ref{fig:text_pca} also demonstates that we are able to capture morphological outliers in the data for the ceramic microspheres and round sand samples. This suggests that the automatic, computationally efficient nature of this method, coupled with its statistical power, would make it amenable to the development of statistical process control techniques for the continuous improvement and quality control of processes in which shape is a key process parameter.

These results reflect the intuition around the shape of these three samples gained from the photographs in Figure \ref{fig:samples}. Ceramic microspheres are specifically manufactured to have consistent morphology, which is reflected in both the mean distance to the perfect circle and in the tight grouping of the material, but at the expense of a higher production cost. The round sand represents a high-quality pure quartz sand that has consistent morphology often obtained from sandstone, representing a more affordable alternative to man-made materials. The angular sand is often not used in applications where the sand requires homogeneous physical properties. However this can be very useful in applications where resistance to shear and cushioning of the sand can be beneficial, as is needed in equestrian arenas to reduce potential injuries to horses \cite{stavermann2014equestrian}. 

\begin{table}[ht]
\begin{center}
\begin{tabular}{ccccc}
  \hline
 & Ceramic & Round & Angular  \\ 
  \hline
Ceramic & - &  &  &  \\ 
  Round & 0.045 & - &  &  \\ 
  Angular & $6.84e^{-9}$ & $2.40e^{-7}$ & - &  \\ 
   \hline
\end{tabular}
\caption{t-test p-values between the different samples.}
\label{tbl:1}
\end{center}
\end{table}

\section{Discussion and Conclusion}

We have presented a Riemannian geometric framework for the automatic quantification of shape for materials. We show that shapes can be represented as points on a non-linear Riemannian manifold. We can exploit the geometric and mathematical properties of the Riemannian manifold to construct a measure of the difference in shape between materials, known as shape space analysis. Riemannian manifolds are equipped with a tangent space defined at each point of the manifold, allowing us to project data from the non-linear manifold to a linear (vector) space. This allows us to directly perform data analysis such as hypothesis testing and dimensionality reduction which we demonstrate in a test on granular material samples from Covia Corp. In our case study we show that the Riemannian geometric framework can be applied automatically in the analysis of granular material images. We show that this method provides generalizable quantification of the differences in the shape of  the individual particles, which is in contrast with the subjective, manual, and bespoke methods used commonly in the industry, such as the Krumbien-Sloss chart method. 

In future work we look to integrate this framework into the statistical quality control of the shape of materials. This can be applied in the mining/manufacturing of granular materials to control the overall quality of the material, in the production of pharmaceuticals where the materials morphology can dramatically impact the solubility and effectiveness of the material, in the production of 3-dimensional printed materials, and in the manufacturing of biomaterials such as cells or fungi where geometry of the material is a key process parameter \cite{quintanilla2015fungal,wu2018quality, de2005real,zheng2016frac, braga2022relevance}. We will also explore the use of the geometric shape of a material and statistics around its morphology to predict bulk physical properties, such as porosity, hardness, and solubility. To accomplish this we will also integrate these geometric methods with \emph{topological} methods derived from the field of Topological Data Analysis to provide a holistic, complementary analysis of a material's morphology \cite{smith2021topological, smith2021euler}. Furthermore, the shape of a material lies on a differentiable manifold on which we can compute gradients and perform optimization \cite{ring2012optimization}. We look to leverage this structure to develop methods in which we can change a material's shape to perform multi-scale optimization and design of materials and processes. 

\bibliography{References}

\end{document}